# On Apparent Absence of "Green Gap" in InGaN/GaN Quantum Disks and Wells Grown by Plasma-Assisted Molecular Beam Epitaxy

Sharif Md. Sadaf[1], Nirmal Anand[1], Emile A. Carbone[1], Dipon K. Ghosh[1], Haipeng Tang[2]

[1]*Centre Energie, Matériaux et Télécommunications, Institut National de la Recherche Scientifique (INRS)-Université du Québec, 1650 Boulevard Lionel-Boulet, Varennes, Quebec J3X 1P7, Canada.*

[2]*Advanced Electronics and Photonics, National Research Council Canada, Ottawa, K1A 0R6, Canada.*

[*]E-mail : sharif.sadaf@inrs.ca

**ABSTRACT**

III-nitride based full-color blue, green and red-light emitting diodes are critically important for a broad range of important applications. To date, however, green or red color III-nitride light emitters grown by conventional growth techniques are limited in efficiency compared to blue emitters. As opposed to metal-organic chemical vapor deposition (MOCVD), while grown by plasma-assisted molecular beam epitaxy (PAMBE), the most intense emission is generally observed in the green spectral region in InGaN/GaN based light emitters. Such counterintuitive phenomenon of efficiency increase with increasing emission wavelength has been observed in both InGaN/GaN quantum-disks in nanowire and planar quantum-wells structures grown by PAMBE. Here, we experimentally show that the apparent absence of "green gap" in longer green wavelength is due to the difficulty of elimination of indium-rich non-radiative clusters and phase segregation in shorter blue wavelength quantum wells/disks. Excess indium due to the dissociation of the In-N bonds during growth lead to nitrogen vacancies and metallic inclusions. In radio-frequency PAMBE, the energy of the nitrogen radicals was found to be a driving force for indium incorporation. Our detailed growth and associated photoluminescence studies suggests that uniform phase and absence of metallic inclusion is the underlying mechanism of efficient green InGaN/GaN quantum wells/disks grown with sufficiently energetic plasma flux. Our study is valid for achieving very efficient green and red color InGaN/GaN and breaking the "green gap" bottleneck in quantum wells/disks grown by state-of-the-art high-power plasma-assisted molecular beam epitaxy.

# 1. Introduction

High efficiency InGaN/GaN quantum disks (QD) and quantum well (QW) based light-emitting diodes (LEDs) and laser diodes (LDs) operating in the green and red spectral ranges are in demand for full-color solid state lighting technology and display applications.[1, 2,29] The current full-color lighting technology relies on inefficient and expensive phosphor down conversion process to achieve full-color lights from blue light emitters. Therefore, direct color mixing of red, green and blue light emitters is highly desired to realize full-color white LEDs.[1, 3] However, the best reported external quantum efficiency (EQE) in the green/yellow spectral wavelengths is <30%, whereas the EQE in the blue spectral range is ~80%.[4-6] Such low efficiency in the green spectral region is called "green gap", and is considered to be the critical constraint for phosphor-free, pure semiconductor based white color light emission.[3] Interestingly, despite the huge Stokes loss during phosphor conversion, phosphor converted InGaN/GaN white LEDs show higher efficiency than green or red LEDs.[3] The performance of III-nitride deep green emitters has been severely limited by the large lattice mismatch between InN and GaN, the large solid-solubility or miscibility gap and the quantum confined Stark effect (QCSE).[3, 7-9]. To date, it has remained challenging to incorporate sufficient amount of indiums (In) into the InGaN QWs/QDs without defect because of In segregation and/or phase separation. Furthermore, In incorporation largely depends on the epitaxial material growth techniques and growth parameters.[10-13]

To date, plasma-assisted molecular beam epitaxy (PAMBE) and metal organic chemical vapor deposition (MOCVD) have remained reliable growth techniques for realizing high quality, defects free InGaN/GaN QWs/QDs based LEDs/LDs.[12, 14] A comparative study experimentally demonstrated that while InGaN/GaN planar QWs grown by MOCVD show the brightest emission in the blue spectral region, InGaN/GaN QDs-in-nanowire grown by PAMBE show most efficient emission in the green spectral region.[15] Previous reports also demonstrated that there is a monotonic decrease in efficiency with the decrease in In content or vice-versa in QDs-in-nanowire grown by PAMBE.[8, 16-18] The complicated interplay of surface potentials and fields associated with large spontaneous/piezoelectric polarizations were attributed to such unusual behavior.[8, 15] Although ammonia based MBE and MOCVD is a mature technology for growing high efficiency blue InGaN/GaN QWs LEDs, the plasma-assisted MBE growth technique offers several unique advantages including high quality material, high In-content InGaN growth and opportunities for controllable compositional tuning.[9, 11-13, 19] For III-nitride materials growth, the growth windows and growth mechanisms are very different in PAMBE as opposed to MOCVD.[20, 21] Unlike MOCVD, PAMBE offers the unique advantage of growing at low temperature under *non-equilibrium* conditions and the supply of active nitrogen is not dependent on the substrate growth temperature.[10, 11] High In content InGaN growth with superior crystalline quality at significantly lower growth temperature was previously demonstrated by metal-modulated PAMBE.[9, 19] The technique allows easier growth of high indium composition InGaN and high efficiency *p*-type doping without post-growth annealing.[22, 23]

Previous studies suggest that both catalyst-free InGaN/GaN QDs-in-nanowire and planar quantum wells can be spontaneously grown by PAMBE. To date, however, there have been few reports on detailed growth studies of InGaN/GaN QWs and QDs-in-nanowire grown by PAMBE. Moreover, there have been no side-by-side comparison of the InGaN/GaN QWs/QDs grown solely by PAMBE.

In this context, we have systematically investigated the design, molecular beam epitaxy growth and photoemission characteristics of InGaN/GaN emitters (QWs/QDs) both in planar and nanowire geometries operating in the visible spectral range. Plasma-assisted MBE grown, nearly defect-free nanowire and planar InGaN/GaN emitters show maximum efficiency in the green spectral region which is not commonly observed by MOCVD or ammonia MBE growth. Our detailed studies suggest that maximum achievable emission efficiency in the blue spectral region is critically limited by RF plasma power and/or active nitrogen species, Ga/In ($\Phi_{Ga}/\Phi_{In}$) fluxes and growth temperatures.

In this work, we have studied two different types of InGaN emitters, including InGaN/GaN QDs-in-nanowire on Si and planar QWs on sapphire heterostructures by plasma-assisted MBE. The growth conditions were carefully optimized to get the highest overall luminous efficiency for both classes of light emitters. Here, we demonstrate the critical role of active nitrogen flux, In/Ga fluxes, and growth temperature on the overall emission efficiency of InGaN/GaN QDs and QWs. Our detailed study pin-pointed the causes of the apparent absence of the "green gap" in PAMBE grown light emitters.

## 2. Experimental methods:

Schematically shown in Fig. 1, all the samples were grown in a SVTA nitride MBE system equipped with a Veeco unibulb plasma source. For the quantum-disks-in nanowire samples, the substrate surface oxide was desorbed *in-situ* at 870°C. First, an unintentionally doped GaN nanowire contact layer was directly grown on *n*-Si (111) substrate at a growth temperature of 760°C, nitrogen flow rate of 1.0 standard cubic centimeter per minute (sccm), forward plasma power of 350 W, and Ga beam equivalent pressure of $6\times10^{-8}$ Torr. Subsequently, 5 N-polar self-organized InGaN/GaN disks-in-nanowires were grown. The InGaN/GaN quantum disks were grown at relatively low temperatures (∼650°C) to enhance the In incorporation into the disks. Each ∼3 nm InGaN quantum disk was subsequently capped by a GaN layer of ∼3 nm. The growth conditions, including metal fluxes of In and Ga ($\Phi_{In}/\Phi_{Ga}$), growth temperature and $N_2$ flow were optimized. The base pressures of the growth chamber were $10^{-10}$ Torr under idle condition and $2\times10^{-5}$ Torr during growth due to active N*. The impinging metal fluxes were measured *in-situ* by a flux gauge, and Ga beam equivalent pressure (BEP) of $1.4\times10^{-7}$ Torr was maintained for the contact layer. For the active QDs region, Ga BEP of $7.6\times10^{-8}$ Torr was maintained and the In BEP was varied within the ranges of $7.7\times10^{-8}$ Torr to $2.0\times10^{-7}$ Torr.[1] All the nanowires are of wurtzite crystal structure and possess N-polarity.

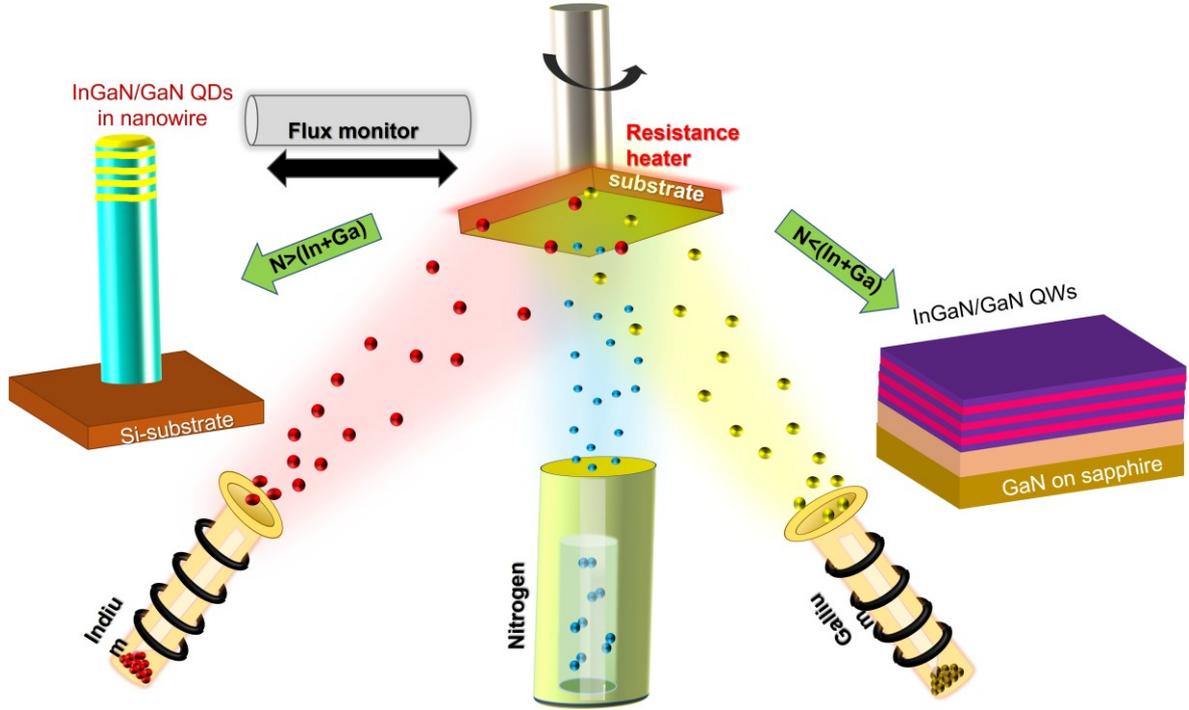

**Fig. 1.** (a) Schematic illustration of N face polar InGaN/GaN QDs-in nanowires and metal face polar InGaN/GaN QWs grown on Si and sapphire substrate by radio frequency plasma assisted molecular beam epitaxy.

For the quantum well growths, 4 μm thick unintentionally doped HVPE GaN on sapphire templates were used as substrates, the same plasma source and growth chamber were used. The surface native oxides of the substrates were removed by *ex-situ* HF dipping and *in-situ* Ga adsorption/desorption thermal treatment at 720°C. Prior to InGaN/GaN QWs growth, a 0.4 μm thick GaN buffer layer was grown under intermediate Ga-rich conditions, to get rid of possible impurities for the subsequent growth. Subsequently, 5 Ga-polar planar InGaN/GaN (3 nm/7 nm) QWs growths were performed at ~590°C. During the QWs growth, excess In was desorbed from the growth surface into vacuum. Nitrogen flow rate of 1.25 standard cubic centimeter per minute (sccm) and forward plasma power of 220W-270W were used. The In/Ga metal fluxes $\Phi_{In}/\Phi_{Ga}$, growth temperature and radio-frequency plasma powers were systematically varied to tune the emission wavelength and enhance the luminous efficiency.

The actual growth temperatures on different substrates were measured using a calibrated pyrometer. The pyrometer calibration was done from 1x1 to 7x7 transition using reflection high energy electron diffraction (RHEED) of Si (111) substrates. The growth temperatures were further calibrated *in-situ* by an optical reflectance measurement system by monitoring Ga desorption time constants from Si, GaN and GaN-on sapphire substrates. Such growth temperature calibration technique can reliably determine the real growth temperature since excess Ga deposition on GaN and desorption from the GaN surface depend on the growth temperatures and surface roughness. It is worthwhile mentioning that a few nm GaN was deposited on the bare substrates (Si, GaN or Sapphire) prior to these Ga sticking and reflectance tests. By depositing a liquid Ga bilayer and monitoring the reflectance rise and decay characteristics, we derived the time constants for desorption of the deposited Ga atoms from different substrates, as illustrated in Figure 2a. At low temperatures, the reflectance reaches saturation upon the Ga deposition, and then decays slowly after the Ga shutter is closed. The reflectance intensity saturation indicates a full Ga bilayer is established. At higher temperatures, unlike low temperatures, reflectance intensity does not reach saturation point and the Ga desorption rate increases (decay time constants are shortened). A quantitative and reproducible correlation was found between the Ga desorption time constant and the substrate temperature as shown in Fig. 2b. Depending on the Ga desorption time constant from the GaN growth front, we divided the growth regimes (both for QDs and QWs) into three regimes including, a) *low temperature* b) *intermediate temperature* and c) *high temperature*. It is worthwhile mentioning that different segments including p-/n-GaN nanowire, InGaN/GaN QDs and QWs were grown at different growth temperatures from low-T to high-T. Accurate growth temperature determination was instrumental for our study since InGaN/GaN active region growths and associated luminescence critically depend on the real growth temperature. Similar calibration of substrate temperature by Ga desorption time constant has been reported before, except RHEED was used to monitor the Ga adsorption/desorption characteristics. Subsequently, after growth, structural properties of InGaN/GaN quantum disks-in-nanowire LEDs were characterized by field emission scanning electron microscopy (SEM). The nanowires are vertically aligned on the Si substrate with diameters and densities in the range of 50 nm−100 nm and $1 \times 10^9$ cm$^{-2}$, respectively (not shown).

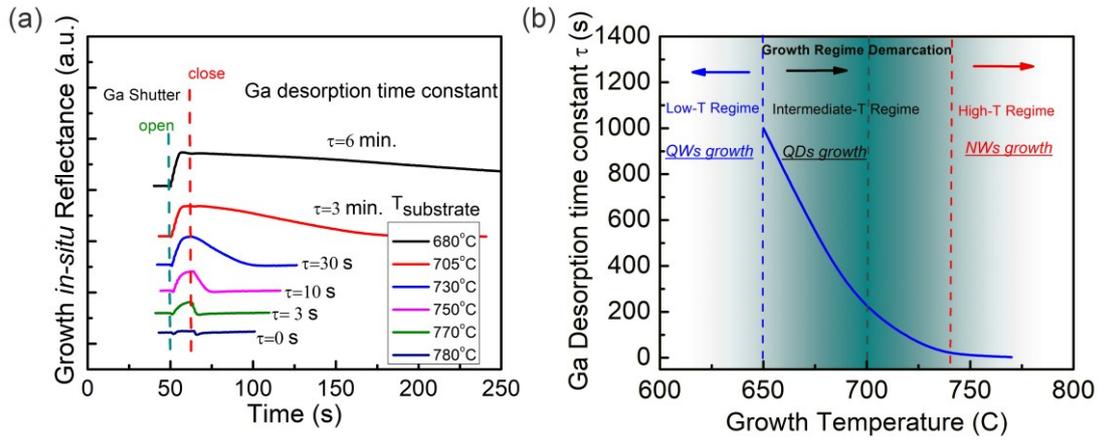

**Fig. 2:** (a) *in-situ* reflectance response to deposition of two monolayers of Ga and its subsequent desorption at different temperatures. (b) Ga desorption time constant versus substrate temperature and demarcation of temperature regimes for planar QWs, QDs-in nanowire and nanowire growths.

## 3. Results:

Shown in Fig. 3a is the normalized photoluminescence (PL) characteristic measured at room temperature with a 325 nm laser excitation. A single PL emission peak centered at ~555 nm was measured that corresponds to emission from the quantum disk active region. Shown in the inset, the power dependent PL emission tuning toward shorter wavelength further confirms the presence of InGaN/GaN quantum disks. It also suggests that the emission peak at ~555 nm at low excitation power (~513 nm at high excitation power) is not related to defect induced yellow luminescence and indeed comes from the InGaN/GaN QDs. The observed blue shift (~42 nm) can be attributed to the band filling effect related to increased photogenerated carrier recombination by higher power laser excitation.[24,25] The InGaN/GaN
8

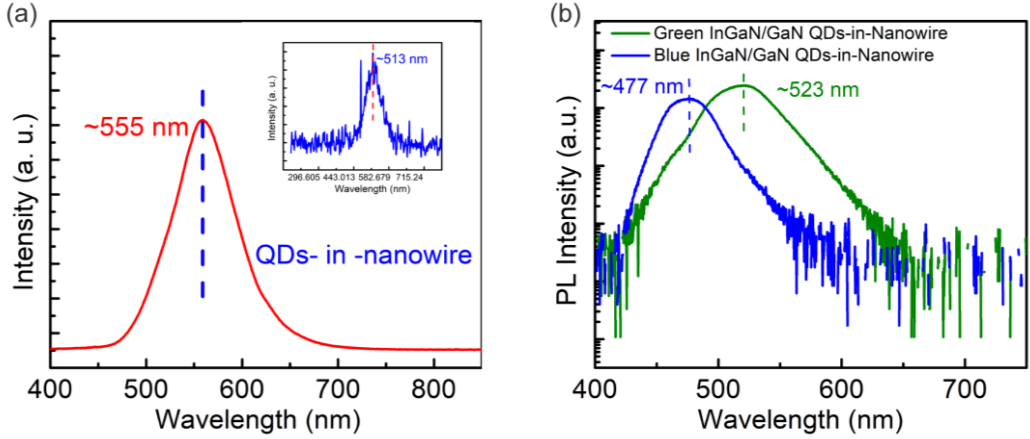

**Fig. 3.** (a) PL spectra of InGaN/GaN QDs in nanowire measured at room temperature, inset shows the PL emission tuning. (b) PL spectra of InGaN/GaN dot-in-nanowire LEDs grown on Si substrate. The LEDs were grown under identical plasma power ~350 W, the growth temperature and In/Ga fluxes were optimized for the color tuning.

QDs peak position corresponds to In content of 23% which is approximately determined from the equation $E_g(x) = xE_g,\text{InN} + (1-x)E_g,\text{GaN} - bx(1-x)$, where $x$ is the In composition, $E_g$ is the bandgap energy, and b is bowing parameter that is generally in the range of 1-6 eV.[7] In the present work, b is taken to be 3.6 eV, $E_g$ (GaN) and $E_g$ (InN) are 3.4 and 0.7 eV.[7]

Contrary to InGaN/GaN QWs, previous reports on InGaN/GaN disks-in nanowire LEDs suggest that green color emission can be relatively easily achieved by PAMBE growth technique and the best optical efficiency can only be achieved in the green spectral range.[14, 15, 26, 27] While this could be interpreted as an apparent absence of the "green gap" in disks-in-nanowire heterostructures, these observations can be attributed to several factors. First, the large surface-to-volume ratio and associated strain relaxation in the nanowire sidewalls facilitates higher In incorporation without the formation of extended defects or material quality degradation[3], which is difficult or impossible in planar InGaN/GaN QWs grown by MOCVD or ammonia MBE.[8, 15] Such strain relaxation also helps minimize quantum-confined Stark effect (QCSE) associated wavelength shift.[26, 28] Second, the spontaneous bottom up nanowires are generally grown under N-rich condition with significantly higher radio frequency (rf) plasma power ~350-400 W that leads to significantly high N-flux and/or active N species. By using such high N-flux ($\Phi_N^* \gg \Phi_G$), In incorporation can be significantly enhanced even at growth temperatures >650°C. Turski *et al.* have demonstrated that by supplying higher N-flux, it is indeed possible to increase In composition as high as 10%.[29] Third, In incorporation can be further enhanced in the QDs by changing the shape and size of the QDs.[30] Plasma assisted MBE nanowire growth typically takes place in N-rich condition at much higher growth temperatures to enable defect-free one-dimensional growth without coalescence or compact layer formation.[31] Under such N-rich condition, all the impinging In flux can be effectively nitrided by the presence of active N* in the growth front. Additionally, high plasma power kinetically favors more In incorporation into the QDs and reduces In-rich cluster formation. Fourth, because of the QDs-in-nanowire geometry and adatom diffusion kinetics at high growth temperature in N-rich conditions, the nanowires generally show Nitrogen (N) face polarity; previous study also suggested that it is easier to incorporate more In into N-face InGaN layers.[22] Moreover, we found that, at >650°C growth temperature, In/Ga-N decomposition can be effectively suppressed by supplying higher active N* species and/or plasma flux. These factors, by and large, explain efficient green emission in plasma assisted MBE grown nanowires and uniform phase formation with lower defect densities. On the other hand, low In content blue emission is somewhat difficult because of the high plasma power/energetic nitrogen species and strain relaxation on the sidewalls. Prior studies primarily focused on the complex interplay of strain to explain this apparent absence of "green gap" in InGaN nanowires.[8] Previously it was suggested that by decreasing or increasing growth temperatures, In incorporation and/or emission color can be controllably tuned in the QDs.[32] Albeit, it is possible to tune the emission color just by changing the growth temperature, however, the best emission intensity remains in the green spectral range in all the previous reports on QDs-in-nanowires grown by PAMBE.[1, 30, 33] Contrary to previous studies on QDs-in-nanowires, our study suggests that there is a narrow growth window for blue color emission, and can be achieved by carefully controlling the In-flux, growth temperature and plasma power (shown in Fig. 3b). Contrary to prior PAMBE arts, tuning emission wavelength and intensity by simply increasing or decreasing growth temperature is not viable rather this technique will result in more phase segregation and inhomogeneous broadening in InGaN/GaN QDs. As shown in Fig. 3b, the emission intensity of blue light exhibits less intense emission compared to the green light emission while keeping the same plasma power and $\Phi_N/\Phi_{Ga}$ ratio. In addition to strain, critical growth parameters and energetic plasma are the underlying causes of such apparent absence of the "green gap" in the QDs-in-nanowire heterostructures.

In contrast to QD-in-nanowire growth, planar Metal (M)-face polarity InGaN/GaN quantum wells growth is generally done in a very different growth regime (III/V>1, metal-rich). For our comparative study, 5 InGaN/GaN quantum wells were grown on GaN (M-face polar) on sapphire template in a broad range of growth conditions varying the plasma power and metal fluxes. Here, we have studied two different sets of samples with different plasma power and In flux. In the first set of samples, we studied four samples varying plasma power and keeping In flux same (820ºC), the samples are denoted as QW-A (270 W), QW-B (250 W), QW-C (220 W) and QW-D (200 W). In the second set of samples, we studied three samples including sample QW-A1 (270 W and 820ºC), QW-B1 (250 W and 810ºC) and QW-C1 (240W and 800ºC) by varying both plasma power and In flux to get the brightest emission in both blue and green wavelengths. For planar QWs, lower plasma power was used, and the barrier growth was done under slightly N-rich, metal-rich and intermediate metal-rich conditions ($\Phi_N/\Phi_G\sim1$). Intermediate metal-rich ($\Phi_G+\Phi_{In}>\Phi_N$) growth condition was used for the growth of QWs, where there is an excess of In adlayer. InGaN/GaN quantum-wells growths temperatures were well below those needed for the fast desorption of excess In atoms. In these growth conditions, excess In due to In-N bond dissociation can lead to nitrogen vacancies and metallic inclusions. In contrast to QDs-in-nanowire, the strain could be higher in the planar InGaN/GaN QWs that results in quantum confined Stark effect, making[34] indium incorporation more difficult. To minimize defects, In incorporation was tuned by changing the plasma power rather by lowering growth temperature (Shown in Fig. 4a). In this set of samples, the brightest PL was obtained at green spectral range (QW- A, peak posiiton~500 nm) when using high plasma power and high indium flux. It is possible to tune the emission wavelength towards blue region by reducing the plasma power and/or the indium flux. Shown in Fig. 4a, when only the plasma power was reduced while keeping the same high indium flux, the emission was shifted to shorter wavelengths (QW-B, -C and -D), but with dramatically reduced emission efficiency. Table-1 summarizes the details of the growths of QW-A,-B,-C and -D. We believe the formation of non-radiative excess indium-rich clusters/precipitates are responsible for the weak blue/violet emissions. Interestingly, by decreasing the plasma power and the indium flux simultaneously, such indium-rich clusters could be reduced, leading to enhanced emission in the blue spectral region (QW-B1, at ~460 nm) as shown in Fig. 4b. Nonetheless, the emission intensity is still the most efficient in the green spectral range (QW-A1, at ~500 nm) in this set of samples. We have shown that in addition to the $\Phi_{In}/\Phi_{Ga}$ fluxes and growth temperatures, the energy distribution of the active N$^*$ species as characterized by the plasma power plays an important role in tuning the wavelength from green to blue or vice-versa. In the following section, a correlative interpretation is proposed to explain why it has been seemingly easier to achieve high luminescence efficiency in the green region rather than in the shorter wavelength region with the plasma-assisted MBE technique.

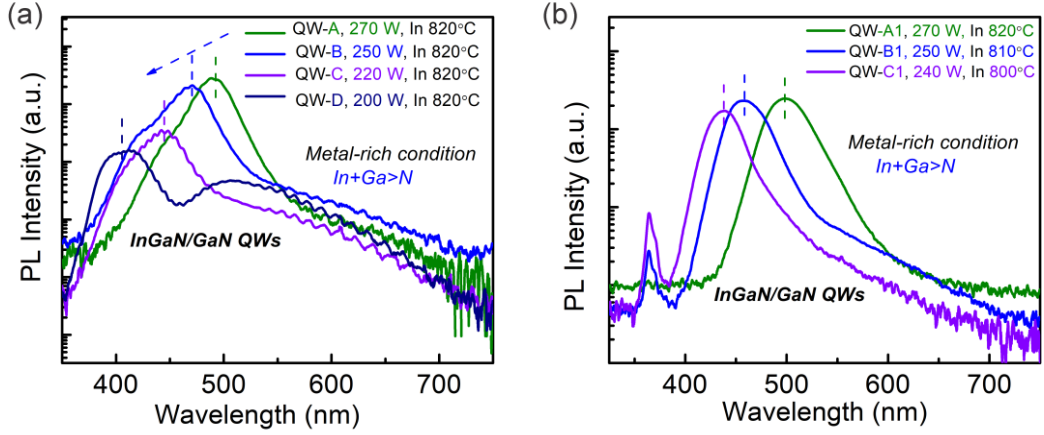

**Fig. 4.** (a) Schematic illustration of InGaN/GaN QWs grown GaN-on Sapphire substrate. PL spectra for planar InGaN/GaN quantum wells grown with (a) identical indium flux and varied plasma power (QW-A,-B,-C and-D) , (b) varied indium flux and varied plasma power (QW-A1,-B1 and-C1).

Table-1. Growth conditions and details of the PAMBE-grown of QW-A,-B,-C and -D.

| Samples# | Plasma Power (W) | Wavelength (nm) | In composition (%) | Growth temperature | In-flux (BEP in Torr), at 820°C |
|---|---|---|---|---|---|
| QW-A | 270 | 490 | 19 | 590°C | $2.5\times10^{-7}$ |
| QW-B | 250 | 460 | 15 | 590 °C | $2.5\times10^{-7}$ |
| QW-C | 220 | 440 | 12 | 590 °C | $2.5\times10^{-7}$ |
| QW-D | 200 | 410 | 6 | 590 °C | $2.5\times10^{-7}$ |

## 4. Discussion:

Phase separation is a general phenomenon in InGaN/GaN QWs or QDs due to the lattice mismatch and miscibility gap between InN and GaN.[2] Previous reports suggest that decrease in radiative recombination in high In-content InGaN/GaN QWs is due to random alloy fluctuation.[3, 35] High growth temperature and unbalanced III/V ratio can be a possible driving force for inducing InGaN phase segregation.[11] However, in PAMBE, due to distribution of radical energies (e.g. atomic nitrogen versus excited molecules) in the plasma, phase segregation can be triggered in certain situations. Our study suggests that, with sufficiently energetic plasma flux, metallic inclusions free and uniform InGaN phases can be realized. For blue QWs growth, if the growth temperature and $\Phi_{In}/\Phi_{N}^*$ fluxes are not properly optimized, phase separation could occur forming indium-rich clusters. Shown in Fig. 5a, a shoulder peak related to higher indium composition clusters appears at longer wavelength (at ~490 nm) than the targeted blue emission peak (~400 nm) in the PL spectrum. Such higher In-content clusters are partially nitrided and contain numerous non-radiative centers.

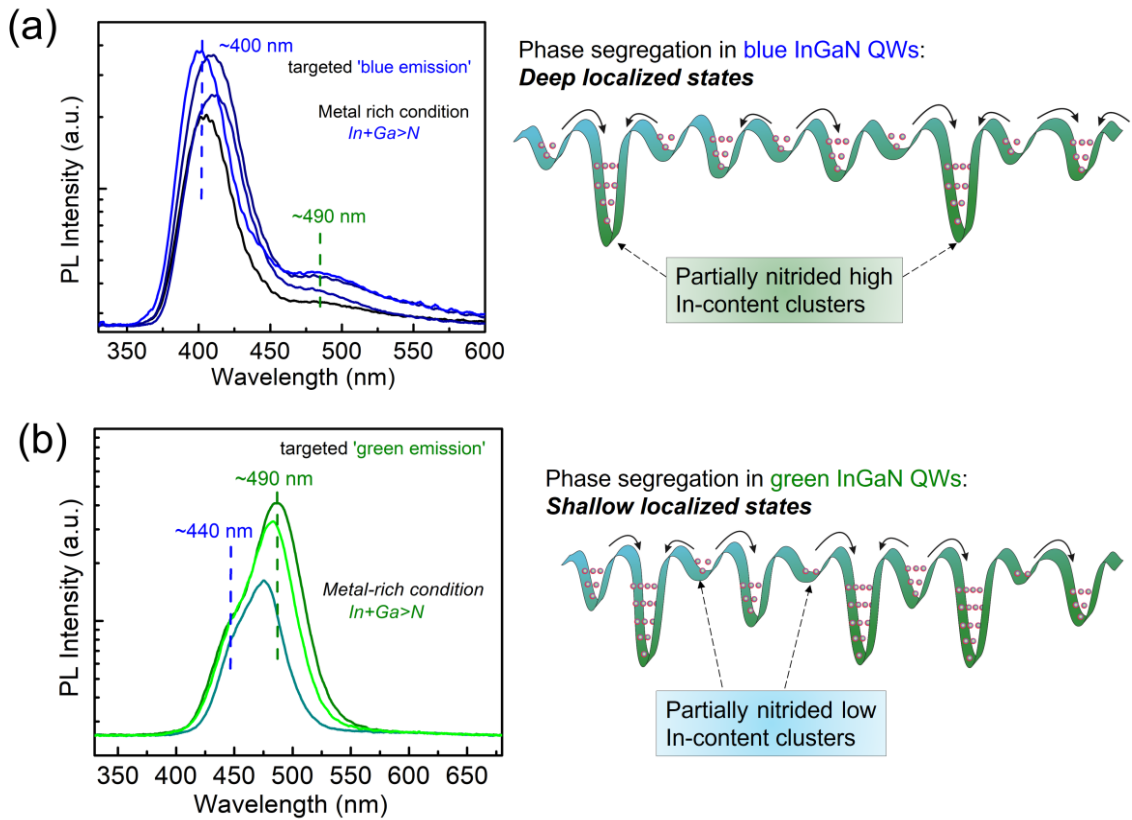

**Fig. 5.** Schematic band diagrams indicating possible carrier recombination dynamics at both shallow and deep localized states when phase separation happens in PAMBE growth. (a) InGaN phase separation related carrier recombination dynamics when the 'blue emission' was targeted but the critical growth parameters are not properly tuned (b) InGaN phase separation related carrier recombination dynamics when the 'green emission' was targeted but the critical growth parameters are not properly tuned.

It is worth pointing out that such discretely located high In-content clusters are at lower energy levels (deep localized states) than the uniform quantum well region. As such, these lower energy clusters capture carriers relatively easily as illustrated in the band diagram in Fig. 5a. Such partially nitrided In-rich phases and non-radiative centers can dramatically reduce the targeted blue emission's efficiency. Carriers captured by the clusters recombine both radiatively (at longer wavelengths) and non-radiatively, thereby reducing the overall efficiency of the blue emitter. As opposed to blue QWs, green QWs are generally grown at lower temperature and with sufficiently high plasma power which help reduce In phase separation. Yet, there could be certain a degree of phase segregation, however, in this case, the discrete clusters are at higher energy levels (shallow localized states) than the targeted, uniform green emitting QWs, as illustrated by the band diagram in Fig 5b. The lower energy (deep localized states) green quantum wells can capture carriers more effectively than the higher energy discrete blue cluster sites as illustrated in Fig. 5b. In addition, at room temperature, some carriers tend to escape from the shallow localized states with thermal energy (~$K_BT$) and relax back to energetically favorable deep localized states. This suggests that the luminescence efficiency of the green QWs is less sensitive to adverse effects of phase segregation than blue QWs. This phenomenon offers a possible explanation of the apparent absence of "green gap" in PAMBE grown InGaN/GaN QWs and/or QDs if all the PAMBE growth parameters are not properly adjusted.

## 5. Conclusions

In summary, our detailed comparison results of planar Ga-polar InGaN/GaN QWs and N-polar InGaN/GaN QDs-in-nanowire grown by PAMBE provides a new paradigm for full-color compositional tuning. Detailed side by side comparison suggests that maximum achievable efficiency in blue/green QWs and QDs critically depend not only on the metal fluxes ($\Phi_{In}/\Phi_{Ga}$) or growth temperature but also on the energetic plasma flux and active N species. To achieve bright blue color emission, plasma power, In/Ga fluxes and growth temperatures should be adjusted accordingly. This result suggests that unlike previous reports, plasma power and/or energetic nitrogen species play a very important role for the defect-free efficient InGaN/GaN light emitters growth. It is observed that, using radio-frequency plasma, InGaN/GaN QDs-in-nanowire with >25%, In can be achieved due to the use of high N-flux and growth temperatures, whereas to get similar amount of In in the InGaN/GaN QWs, N-flux, Ga/In flux and growth temperatures need to be carefully adjusted. Our studies further suggest that PAMBE can be a viable platform for realizing bright phosphor free tunable white color emission. In addition, PAMBE could circumvent green gap and phase separation related problems in the longer wavelength light emitters.

## Acknowledgments

This work was supported by the Natural Sciences and Engineering Research Council of Canada (NSERC) through Discovery and Quantum Alliance Grant programs, MEIE Photonique Quantique Quebec (PQ2) program, and the Canada Research Chair program. We would like toacknowledge CMC Microsystems for the provision of productsand services and fabrication fund assistance using the facilitiesof Laboratory of Mico and Nanofabrication (LMN) at INRS. The authors gratefully acknowledge Ryszard Dabkowski for technical assistance with the MBE growth.

**References:**

(1) Nguyen, H. P.; Cui, K.; Zhang, S.; Fathololoumi, S.; Mi, Z. Full-color InGaN/GaN dot-in-a-wire light emitting diodes on silicon. *Nanotechnology* **2011**, *22* (44), 445202.

(2) Wang, Q.; Gao, X.; Xu, Y.; Leng, J. Carrier localization in strong phase-separated InGaN/GaN multiple-quantum-well dual-wavelength LEDs. *Journal of Alloys and Compounds* **2017**, *726*, 460-465.

(3) Auf der Maur, M.; Pecchia, A.; Penazzi, G.; Rodrigues, W.; Di Carlo, A. Efficiency Drop in Green InGaN/GaN Light Emitting Diodes: The Role of Random Alloy Fluctuations. *Phys Rev Lett* **2016**, *116* (2), 027401.

(4) Humphreys, C. J.; Griffiths, J. T.; Tang, F.; Oehler, F.; Findlay, S. D.; Zheng, C.; Etheridge, J.; Martin, T. L.; Bagot, P. A. J.; Moody, M. P.; et al. The atomic structure of polar and non-polar InGaN quantum wells and the green gap problem. *Ultramicroscopy* **2017**, *176*, 93-98.

(5) Liu, J.-L.; Zhang, J.-L.; Wang, G.-X.; Mo, C.-L.; Xu, L.-Q.; Ding, J.; Quan, Z.-J.; Wang, X.-L.; Pan, S.; Zheng, C.-D.; et al. Status of GaN-based green light-emitting diodes. *Chinese Physics B* **2015**, *24* (6).

(6) Karpov, S. Y. Carrier localization in InGaN by composition fluctuations: implication to the "green gap". *Photonics Research* **2017**, *5* (2).

(7) Yam, F. K.; Hassan, Z. InGaN: An overview of the growth kinetics, physical properties and emission mechanisms. *Superlattices and Microstructures* **2008**, *43* (1), 1-23.

(8) Marquardt, O.; Hauswald, C.; Wolz, M.; Geelhaar, L.; Brandt, O. Luminous efficiency of axial In(x)Ga(1-x)N/GaN nanowire heterostructures: interplay of polarization and surface potentials. *Nano Lett* **2013**, *13* (7), 3298-3304.

(9) Fischer, A. M.; Wei, Y. O.; Ponce, F. A.; Moseley, M.; Gunning, B.; Doolittle, W. A. Highly luminescent, high-indium-content InGaN film with uniform composition and full misfit-strain relaxation. *Applied Physics Letters* **2013**, *103* (13).

(10) Hu, C.-H.; Lo, I.; Hsu, Y.-C.; Shih, C.-H.; Pang, W.-Y.; Wang, Y.-C.; Lin, Y.-C.; Yang, C.-C.; Tsai, C.-D.; Hsu, G. Z. L. Growth of InGaN/GaN quantum wells with graded InGaN buffer for green-to-yellow light emitters. *Japanese Journal of Applied Physics* **2016**, *55* (8).

(11) Valdueza-Felip, S.; Bellet-Amalric, E.; Núñez-Cascajero, A.; Wang, Y.; Chauvat, M. P.; Ruterana, P.; Pouget, S.; Lorenz, K.; Alves, E.; Monroy, E. High In-content InGaN layers synthesized by plasma-assisted molecular-beam epitaxy: Growth conditions, strain relaxation, and In incorporation kinetics. *Journal of Applied Physics* **2014**, *116* (23).

(12) Gačević, Ž.; Gómez, V. J.; Lepetit, N. G.; Soto Rodríguez, P. E. D.; Bengoechea, A.; Fernández-Garrido,

S.; Nötzel, R.; Calleja, E. A comprehensive diagram to grow (0001)InGaN alloys by molecular beam epitaxy. *Journal of Crystal Growth* **2013**, *364*, 123-127.

(13) Yamaguchi, T.; Uematsu, N.; Araki, T.; Honda, T.; Yoon, E.; Nanishi, Y. Growth of thick InGaN films with entire alloy composition using droplet elimination by radical-beam irradiation. *Journal of Crystal Growth* **2013**, *377*, 123-126.

(14) Frost, T.; Jahangir, S.; Stark, E.; Deshpande, S.; Hazari, A.; Zhao, C.; Ooi, B. S.; Bhattacharya, P. Monolithic electrically injected nanowire array edge-emitting laser on (001) silicon. *Nano Lett* **2014**, *14* (8), 4535-4541.

(15) Feix, F.; Flissikowski, T.; Sabelfeld, K. K.; Kaganer, V. M.; Wölz, M.; Geelhaar, L.; Grahn, H. T.; Brandt, O. Ga-Polar (In,Ga)N/GaN Quantum Wells Versus N-Polar (In,Ga)N Quantum Disks in GaN Nanowires: A Comparative Analysis of Carrier Recombination, Diffusion, and Radiative Efficiency. *Physical Review Applied* **2017**, *8* (1).

(16) Wolz, M.; Lahnemann, J.; Brandt, O.; Kaganer, V. M.; Ramsteiner, M.; Pfuller, C.; Hauswald, C.; Huang, C. N.; Geelhaar, L.; Riechert, H. Correlation between In content and emission wavelength of In(x)Ga(1-x)N/GaN nanowire heterostructures. *Nanotechnology* **2012**, *23* (45), 455203.

(17) Sadaf, S. M.; Ra, Y. H.; Zhao, S.; Szkopek, T.; Mi, Z. Structural and electrical characterization of monolithic core-double shell n-GaN/Al/p-AlGaN nanowire heterostructures grown by molecular beam epitaxy. *Nanoscale* **2019**, *11* (9), 3888-3895.

(18) Cheriton, R.; Sadaf, S. M.; Robichaud, L.; Krich, J. J.; Mi, Z.; Hinzer, K. Two-photon photocurrent in InGaN/GaN nanowire intermediate band solar cells. *Communications Materials* **2020**, *1* (1).

(19) Moseley, M.; Gunning, B.; Greenlee, J.; Lowder, J.; Namkoong, G.; Alan Doolittle, W. Observation and control of the surface kinetics of InGaN for the elimination of phase separation. *Journal of Applied Physics* **2012**, *112* (1).

(20) Skierbiszewski, C.; Turski, H.; Muziol, G.; Siekacz, M.; Sawicka, M.; Cywiński, G.; Wasilewski, Z. R.; Porowski, S. Nitride-based laser diodes grown by plasma-assisted molecular beam epitaxy. *Journal of Physics D: Applied Physics* **2014**, *47* (7).

(21) Chichibu, S. F.; Uedono, A.; Onuma, T.; Haskell, B. A.; Chakraborty, A.; Koyama, T.; Fini, P. T.; Keller, S.; Denbaars, S. P.; Speck, J. S.; et al. Origin of defect-insensitive emission probability in In-containing (Al,In,Ga)N alloy semiconductors. *Nat Mater* **2006**, *5* (10), 810-816.

(22) Akyol, F.; Nath, D. N.; Gür, E.; Park, P. S.; Rajan, S. N-Polar III–Nitride Green (540 nm) Light Emitting Diode. *Japanese Journal of Applied Physics* **2011**, *50* (5R).

(23) Sadaf, S. M.; Tang, H. Mapping the growth of p-type GaN layer under Ga-rich and N-rich conditions at low to high temperatures by plasma-assisted molecular beam epitaxy. *Applied Physics Letters* **2020**, *117* (25).

(24) Schomig, H.; Halm, S.; Forchel, A.; Bacher, G.; Off, J.; Scholz, F. Probing individual localization centers in an InGaN/GaN quantum well. *Phys Rev Lett* **2004**, *92* (10), 106802.

(25) Tangi, M.; Min, J.-W.; Priante, D.; Subedi, R. C.; Anjum, D. H.; Prabaswara, A.; Alfaraj, N.; Liang, J. W.; Shakfa, M. K.; Ng, T. K.; et al. Observation of piezotronic and piezo-phototronic effects in n-InGaN nanowires/Ti grown by molecular beam epitaxy. *Nano Energy* **2018**, *54*, 264-271.

(26) Sadaf, S. M.; Ra, Y. H.; Szkopek, T.; Mi, Z. Monolithically Integrated Metal/Semiconductor Tunnel Junction Nanowire Light-Emitting Diodes. *Nano Lett* **2016**, *16* (2), 1076-1080.

(27) Sadaf, S. M.; Ra, Y. H.; Nguyen, H. P.; Djavid, M.; Mi, Z. Alternating-Current InGaN/GaN Tunnel Junction Nanowire White-Light Emitting Diodes. *Nano Lett* **2015**, *15* (10), 6696-6701.

(28) Anand, N.; Ghosh, D. K.; Abbes, A.; Kundu, M.; Rahman, M. A.; Jenson, C. G.; Morandotti, R.; Baten, M. Z.; Sadaf, S. M. Ultra-dense Green InGaN/GaN Nanoscale Pixels with High Luminescence Stability and Uniformity for Near-Eye Displays. *ACS Nano* **2024**, *18* (39), 26882-26890.

(29) Turski, H.; Feduniewicz-Żmuda, A.; Sawicka, M.; Reszka, A.; Kowalski, B.; Kryśko, M.; Wolny, P.; Smalc-Koziorowska, J.; Siekacz, M.; Muzioł, G.; et al. Nitrogen-rich growth for device quality N-polar InGaN/GaN quantum wells by plasma-assisted MBE. *Journal of Crystal Growth* **2019**, *512*, 208-212.

(30) Ra, Y. H.; Wang, R.; Woo, S. Y.; Djavid, M.; Sadaf, S. M.; Lee, J.; Botton, G. A.; Mi, Z. Full-Color Single Nanowire Pixels for Projection Displays. *Nano Lett* **2016**, *16* (7), 4608-4615.

(31) Nguyen, H. P.; Djavid, M.; Woo, S. Y.; Liu, X.; Connie, A. T.; Sadaf, S.; Wang, Q.; Botton, G. A.; Shih, I.; Mi, Z. Engineering the carrier dynamics of InGaN nanowire white light-emitting diodes by distributed p-AlGaN electron blocking layers. *Sci Rep* **2015**, *5*, 7744.

(32) Gridchin, V. O.; Kotlyar, K. P.; Reznik, R. R.; Dragunova, A. S.; Kryzhanovskaya, N. V.; Lendyashova, V. V.; Kirilenko, D. A.; Soshnikov, I. P.; Shevchuk, D. S.; Cirlin, G. G. Multi-colour light emission from InGaN nanowires monolithically grown on Si substrate by MBE. *Nanotechnology* **2021**, *32* (33).

(33) Wang, R.; Nguyen, H. P.; Connie, A. T.; Lee, J.; Shih, I.; Mi, Z. Color-tunable, phosphor-free InGaN nanowire light-emitting diode arrays monolithically integrated on silicon. *Opt Express* **2014**, *22 Suppl 7*, A1768-1775.

(34) Woo, S. Y.; Gauquelin, N.; Nguyen, H. P.; Mi, Z.; Botton, G. A. Interplay of strain and indium incorporation in InGaN/GaN dot-in-a-wire nanostructures by scanning transmission electron microscopy. *Nanotechnology* **2015**, *26* (34), 344002.

(35) Yang, T.-J.; Shivaraman, R.; Speck, J. S.; Wu, Y.-R. The influence of random indium alloy fluctuations in indium gallium nitride quantum wells on the device behavior. *Journal of Applied Physics* **2014**, *116* (11).